\documentclass[aps,prd,preprint,showpacs,groupedaddress]{revtex4-1}
\usepackage{amssymb}
\usepackage{mathrsfs}
\usepackage{amsmath}
\usepackage{mathtools}
\usepackage{pstricks}
\usepackage{color}
\usepackage{graphicx}
\usepackage{amsthm}

\begin{document}


\title{\bf Critical exponents and amplitude ratios of scalar nonextensive $q$-field theories}



\author{P. R. S. Carvalho}
\email{prscarvalho@ufpi.edu.br}
\affiliation{\it Departamento de F\'\i sica, Universidade Federal do Piau\'\i, 64049-550, Teresina, PI, Brazil}





\begin{abstract}
We compute the radiative quantum corrections to the critical exponents and amplitude ratios for O($N$) $\lambda\phi^{4}$ scalar high energy nonextensive $q$-field theories. We employ the field theoretic renormalization group approach through six methods for evaluating the high energy nonextensive critical exponents up to next-to-leading order while the high energy nonextensive amplitude ratios are computed up to leading level by applying three methods. Later we generalize these high energy nonextensive finite loop order results for any loop level. We find that the high energy nonextensive critical exponents are the same when obtained through all the methods employed. The same fact occurs for the high energy nonextensive amplitude ratios. Furthermore, we show that these high energy nonextensive universal quantities are equal to their low energy extensive counterparts, thus showing that the nonextensivity is broken down at high energies.      
\end{abstract}


\maketitle


\section{Introduction}\label{Introduction} 

\par The Boltzmann-Gibbs theory for describing statistical properties of extensive physical systems has attained a remarkable success \cite{Gibbs}. Despite its triumph in the extensive domain, its generalization to the nonextensive realm was proposed \cite{Tsallis1988}. In the latter case, the nonextensive theory (see Ref. \cite{Tsallis} and references therein) is parametrized by a parameter which characterizes the nonextensivity of the theory, namely the nonextensive parameter $q \in \mathbb{R}$. The extensive theory is recovered in the limit $q \rightarrow 1$. The nonextensive parameter can be used for defining three regimes in which physical systems can be categorized: extensive where $q \rightarrow 1$, nonextensive for $q \neq 1$ but away $q = 1$ and nonextensive for $q \neq 1$ but around $q = 1$. The nonextensive theories for $q \neq 1$ but around $q = 1$ are obtained as the first order Taylor expansion of their nonextensive counterparts for $q \neq 1$ but away $q = 1$ in the region $q \sim 1$.

\par Corresponding quantum $q$-field theories were designed \cite{1674-1137-42-5-053102}. This nonextensive generalization of quantum field theory leads to nonlinear equations \cite{PhysRevLett.106.140601}  and then one needs to consider the physics of nonlinear phenomena, for example the physical behavior of solitons and breathers \cite{A.C.Scott,A.C.Scott2,PolyaninZaitsev,FrenkelKontorova,SulemCSulemP-L}. Such a generalization \cite{PhysRevLett.106.140601} consists in a modification of the terms present in the linear theory by introducing powers characterized by the nonextensive parameter $q$, where the linear theory is recovered in the limit $q \rightarrow 1$ and the extension from one to $d$ dimensions is straightforward. This generalization procedure is opposed to the standard ones in which a nonlinear theory is obtained by just adding nonlinear terms to the linear theory. Furthermore, the extension from one to $d$ dimensions is not straightforward in some situations. As the generalization process leads to nonlinear equations and then to a resulting nonlinear quantum field theory, the superposition principle is no longer valid and the linearity property is lost in that process \cite{1674-1137-42-5-053102}. Also the nonextensive quantum $q$-field $\phi_{q}$ at very high energies is not Lorentz-invariant, thus Lorentz-invariance is lost at very high energies. For a detailed discussion about this subject, see the Ref. \cite{1674-1137-42-5-053102}. Then, the low energy extensive and very high energy nonextensive quantum $q$-field theories are defined for $q \rightarrow 1$ at low energies and for $q \neq 1$ but away $q = 1$ at very high energies, respectively, while the high energy nonextensive ones are defined for $q \neq 1$ but around $q = 1$ at high or intermediate energies. Thus there is a relation between the extensivity and q parameter in quantum field theory, \emph{i. e.} the extensive quantum field theory is obtained from its nonextensive counterpart through the limit $q \rightarrow 1$.

\par The aim of this work is to investigate the nonextensivity of the O($N$) $\lambda\phi^{4}$ scalar high energy nonextensive $q$-field theory through the computation of the all-loop radiative quantum corrections, after a finite next-to-leading order (NLO) evaluation, for dimensions $2<d<4$ through $\epsilon$-expansion techniques in $\epsilon = 4 - d$ to the high energy nonextensive universal critical exponents and amplitude ratios for the referred theory. For that, we apply the field theoretic renormalization group approach \cite{Wilson197475}. In this approach, when the system is undergoing a continuous phase transition, its critical behavior is a result of the fluctuating properties of a fluctuating quantum field $\phi$ whose mean value is associated to the order parameter (magnetization for magnetic systems for example). This field has its values defined at the points of spacetime. These values are correlated and correlation functions can be defined. These correlation functions (one-particle-irreducible ($1$PI) vertex parts) and thermodynamics functions (derived from the effective potential with spontaneous symmetry breaking), near the transition point, present a simple scaling behavior. In this case, the critical scaling behavior of the $1$PI vertex parts and effective potential with spontaneous symmetry breaking is characterized by critical exponents and amplitude ratios, where the latter are obtained as ratios of amplitudes of the thermodynamic functions above and below the transition point. The critical exponents and amplitude ratios can be the same for completely different systems as a fluid and a ferromagnet. When this happens we say that the different systems belong to the same universality class. The critical exponents and amplitude ratios are universal quantities (unlike the amplitudes themselves) and their universal character are related to the universality \cite{PhysRevLett.16.11,JPhysC.2.1883,JPhysC.2.2158,PhysRev.176.738,PhysRevLett.24.1479,Kadanoff} and two-scale-factor universality hypotheses, respectively \cite{PhysRevLett.29.345} where in both cases there are two independent scales, the field and composite field scales (in general, the scales of magnetization and its conjugate field when we are dealing with magnetic systems). These universal quantities do not depend on the microscopic details of the systems as their critical temperatures or form of lattices but on the other hand on their dimension $d$, $N$ and symmetry of some order parameter if the interactions of its constituents are of short- or long-range type. We will be concerned here with the O($N$) universality class. It is reduced to the situations where short-range interactions are present, namely for the Ising ($N=1$), XY ($N=2$), Heisenberg ($N=3$), self-avoiding random walk ($N=0$), spherical ($N \rightarrow \infty$) models etc \cite{Pelissetto2002549}. 

\par The critical exponents and amplitude ratios values can be obtained roughly in the mean field or Landau approximation \cite{Stanley}, where the fluctuations of the fluctuating quantum field are neglected. As near a continuous phase transition the system displays large fluctuations, we have to take into account the fluctuations, which are non-trivially coupled through an effective coupling constant, at all length scales if we need a precise determination of the universal quantities, as opposed to the Landau approximation ($d>4$). The tool capable of attaining that goal is the renormalization group which is the approached here. We then compute the universal quantities in a perturbative expansion in the effective coupling constant or in an equivalent perturbative expansion in the number of loops of which the $1$PI vertex parts and effective potential with spontaneous symmetry breaking are expanded. In the Landau approximation, we compute no loops while as we go further in the perturbation theory in the number of loops, we are evaluating the radiative quantum corrections to the problem considered. Unfortunately, the $1$PI vertex parts and effective potential with spontaneous symmetry breaking are plagued by divergences, thus we have to renormalize them. The former can be computed through six distinct and independent renormalization methods while the latter by applying three different and independent renormalization schemes. As the renormalized theory is attained through the flow of the renormalized coupling constant to the nontrivial solution of the $\beta$-function, the $1$PI vertex parts and effective potential with spontaneous symmetry breaking acquire anomalous dimensions, namely the field and composite field anomalous dimensions. The trivial solution of the $\beta$-function gives the Landau values for the universal quantities while the nontrivial one permit us to compute these quantities containing their radiative quantum corrections. As the critical exponents and amplitude ratios are universal quantities, it does not matter which renormalization method is used for obtaining them and the many renormalization schemes used are useful for checking the final results. As in the field theoretic approach the difference of some arbitrary temperature $T$ and its critical one $T_{c}$ is proportional to the squared mass $m^{2}$ of the fluctuating quantum field, massive and massless theories mean noncritical and critical theories, respectively. Thus the critical exponents must be the same if computed through both massive (in three distinct methods) and massless (through three different schemes) theories, since the critical scaling behavior of the $1$PI vertex parts at and near the transition is the same. On the other hand, some amplitude ratios involve a few amplitudes at the critical temperature, thus requiring the application of just massless (in three distinct methods) theories and not massive ones. Any deviation of the high energy nonextensive universal quantities values from their low energy extensive counterparts will indicate the nonextensivity character of the corresponding theory, at least at high or intermediate energies. In quantitative terms, these universal quantities would must depend on the nonextensive parameter $q$ which would show the nonextensivity of the theory. In fact, the first order Taylor expansion, around $q = 1$, for the nonextensive $q$-Klein-Gordon equation at very high energies was obtained \cite{e19010021} giving the corresponding high energy nonextensive bare free propagator $G_{0B,{q \sim 1}}(k) = \parbox{12mm}{\includegraphics[scale=1.0]{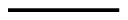}} = 1/(k^{2} + qm_{B}^{2})$ at high or intermediate energies. As it is well known from the original proposition of generalized nonextensive statistical mechanics \cite{Tsallis}, the nonextensive $q$ parameter is a constant. It could be interesting to consider $q$ as a running one, but this case is beyond the purpose of the present work and can be considered in a future work. As there are two independent scales, twelve critical exponents 
\begin{eqnarray}
\alpha, \alpha^{\prime}, \gamma, \gamma^{\prime}, \nu, \nu^{\prime}, \beta, \delta, \eta, \alpha_{c}, \gamma_{c} \nu_{c},
\end{eqnarray}
twelve critical amplitudes
\begin{eqnarray}\label{uhdfuhfuh}
\begin{array}{lcr}
\mbox{\textrm{Critical isochore: $T > T_{c}$, $H = 0$}} &  &  \\
\mbox{} \xi = \xi_{0}^{+}t^{-\nu}, \chi = C^{+}t^{-\gamma}, C_{s} = \frac{A^{+}}{\alpha_{+}}t^{-\alpha} &  \\ [10pt]
\mbox{\textrm{Critical isochore: $T < T_{c}$, $H = 0$}} &  &  \\
\mbox{} \xi = \xi_{0}^{-}t^{-\nu}, \chi = C^{-}t^{-\gamma}, C_{s} = \frac{A^{-}}{\alpha_{-}}t^{-\alpha}, M = B(-t)^{\beta} &  \\ [10pt]
\mbox{\textrm{Critical isotherm: $T = T_{c}$, $H \neq 0$}} &  &  \\ 
\mbox{} \xi = \xi_{0}^{c}|H|^{-\nu_{c}}, \chi = C^{c}|H|^{-\gamma_{c}}, C_{s} = \frac{A^{c}}{\alpha_{c}}|H|^{-\alpha_{c}}, H = DM^{\delta} &  \\ [10pt]
\mbox{\textrm{Critical point: $T = T_{c}$, $H = 0$}} &  &  \\
\mbox{} \chi(p) = \widehat{D}p^{\eta-2} &  \\ \end{array} \nonumber
\end{eqnarray}\label{jlkfdjkglk}
ten relations among the critical exponents
\begin{eqnarray}
\alpha = \alpha^{\prime}, \gamma = \gamma^{\prime}, \nu = \nu^{\prime}, \gamma = \beta(\delta - 1),\alpha = 2 - 2\beta - \gamma \quad
\end{eqnarray} 
\begin{eqnarray}
2 - \alpha = d\nu, \gamma = (2 - \eta)\nu, \alpha_{c} = \frac{\alpha}{\beta\delta}, \gamma_{c} = 1 - \frac{1}{\delta}, \nu_{c} = \frac{\nu}{\beta\delta} \nonumber \\
\end{eqnarray}  
and one relation between the critical amplitudes
\begin{eqnarray}
\delta C^{c}D^{1/\delta} = 1,
\end{eqnarray}
we have to evaluate independently two critical exponents and nine amplitude ratios. Among the high energy nonextensive critical exponents, we have to compute independently $\eta$ and $\nu$ while the nine high energy nonextensive amplitude ratios chosen to be evaluated are that of Ref. \cite{PrivmanHohenbergAharony} which were computed in Refs. therein.

\par This work is organized as follows: Firstly, we have to compute the radiative quantum corrections to the high energy nonextensive critical exponents up to NLO through six distinct and independent renormalization methods for O($N$) $\lambda\phi^{4}$ scalar high energy nonextensive $q$-field theories. Secondly, we have to attain a similar goal but now for high energy nonextensive amplitude ratios up to leading order and by applying three different and independent renormalization schemes. After that, we have to generalize that results for any loop level. At the end, we present our conclusions and perspectives.

\section{high energy nonextensive critical exponents: massive theories}

\par In the following steps, we have to label the high energy nonextensive quantities of interest. For example, we have to set $\eta_{q \sim 1}$ for the $\eta$ exponent and so on. Now we have to compute the high energy nonextensive critical exponents through the methods displayed below.

\subsection{Callan-Symanzik method}

\par For computing the high energy nonextensive critical exponents in the Callan-Symanzik method \cite{Amit}, we need only a minimal set of four Feynman diagrams to evaluate, up to NLO, at fixed dimensionless external momenta at the symmetry point $SP$. This symmetry point is characterized by external momenta fixed at the value $P^{2} = 0$ for low energy extensive systems and the corresponding inverse bare free propagator is given by $k^{2} + 1$ if we use the renormalized mass as a scale unit, where now the momentum $k$ is dimensionless. The observables, being defined at a finite scale, depend on the irrelevant parameters and might well contain important q-dependence. For example, the effective nonextensive mass is given by $qm$. By following the same steps for high energy nonextensive systems, we obtain that the inverse bare free propagator results $k^{2} + q$, where a $q$-dependence is explicit. If we have chosen the effective nonextensive mass $qm$, we would obtain the same bare free propagator as that for the low energy extensive theory and the $q$-dependence of the Feynman diagrams on the nonextensive $q$ parameter would disappear. In fact, the choice of the mass as a scale unit is arbitrary. Even so, we have made a choice such that the $q$-dependence is explicit. This permit us to probe explicitly the effect of $q$ on the high energy nonextensive critical exponents. Then we obtain for the four needed Feynman diagrams evaluated in $d = 4 - \epsilon$    
\begin{eqnarray} 
\parbox{10mm}{\includegraphics[scale=1.0]{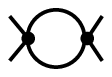}}_{SP} \equiv \parbox{11mm}{\includegraphics[scale=1.0]{fig10.eps}}\Bigg\vert_{P^{2}=0}, 
\end{eqnarray}
\begin{eqnarray}
\parbox{10mm}{\includegraphics[scale=1.0]{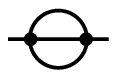}}^{\prime} \equiv \frac{\partial }{\partial P^{2}}\parbox{12mm}{\includegraphics[scale=1.0]{fig6.eps}}\Bigg\vert_{P^{2}=0},
\end{eqnarray}
\begin{eqnarray}
\parbox{12mm}{\includegraphics[scale=.9]{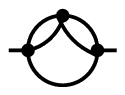}}^{\prime} \equiv \frac{\partial }{\partial P^{2}}\parbox{12mm}{\includegraphics[scale=.9]{fig7.eps}}\Bigg\vert_{P^{2}=0},
\end{eqnarray}
\begin{eqnarray} 
\parbox{13mm}{\includegraphics[scale=.9]{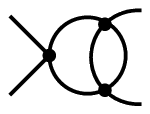}}_{SP} \equiv \parbox{14mm}{\includegraphics[scale=.9]{fig21.eps}}\Bigg\vert_{P^{2}=0}, 
\end{eqnarray}
the following results
\begin{eqnarray}
&&\parbox{10mm}{\includegraphics[scale=1.0]{fig10.eps}}_{SP} = \frac{1}{\epsilon}\left(1 - \frac{1}{2}\epsilon \right)q^{-\epsilon/2},
\end{eqnarray}   
\begin{eqnarray}
&&\parbox{12mm}{\includegraphics[scale=1.0]{fig6.eps}}^{\prime} = -\frac{1}{8\epsilon}\left( 1 - \frac{1}{4}\epsilon +I\epsilon \right)q^{-\epsilon},
\end{eqnarray}  
\begin{eqnarray}
&&\parbox{12mm}{\includegraphics[scale=0.9]{fig7.eps}}^{\prime} = -\frac{1}{6\epsilon^{2}}\left( 1 - \frac{1}{4}\epsilon +\frac{3}{2}I\epsilon \right)q^{-3\epsilon/2},
\end{eqnarray}  
\begin{eqnarray}
&&\parbox{12mm}{\includegraphics[scale=0.8]{fig21.eps}}_{SP} = \frac{1}{2\epsilon^{2}}\left(1 - \frac{1}{2}\epsilon \right)q^{-\epsilon},
\end{eqnarray}  
where the integral $I$ \cite{PhysRevD.8.434}
\begin{eqnarray}
&& I = \int_{0}^{1} \left\{ \frac{1}{1 - x(1 - x)} + \frac{x(1 - x)}{[1 - x(1 - x)]^{2}}\right\}
\end{eqnarray}
is a number and a residual effect of the symmetry point we have chosen. We can now compute the high energy nonextensive $\beta_{q \sim 1}$-function, anomalous dimensions and nontrivial fixed point to obtain 
\begin{eqnarray}\label{fhdfghg}
\beta_{q \sim 1}(u) = -\epsilon u +   \frac{N + 8}{6}\left( 1 - \frac{1}{2}\epsilon \right)q^{-\epsilon/2}u^{2} -  \frac{3N + 14}{12}q^{-\epsilon}u^{3}, 
\end{eqnarray}
\begin{eqnarray}
\gamma_{\phi,q \sim 1} = \frac{N + 2}{72}\left( 1 - \frac{1}{4}\epsilon + I\epsilon \right)q^{-\epsilon}u^{2} -  \frac{(N + 2)(N + 8)}{432}\left( 1 + I \right)q^{-3\epsilon/2}u^{3},  
\end{eqnarray}
\begin{eqnarray}\label{sfdtsvdb}
&&\overline{\gamma}_{\phi^{2},q \sim 1}(u) = \frac{N + 2}{6}\left( 1 - \frac{1}{2}\epsilon \right)q^{-\epsilon/2}u -  \frac{N + 2}{12}q^{-\epsilon}u^{2},
\end{eqnarray}
\begin{eqnarray}
u_{q \sim 1}^{*} = \frac{6\epsilon q^{\epsilon/2}}{(N + 8)}\Bigg\{ 1 +  \epsilon\Bigg[ \frac{3(3N + 14)}{(N + 8)^{2}} + \frac{1}{2} \Bigg]\Bigg\},
\end{eqnarray}
where we have used $\overline{\gamma}_{\phi^{2},q \sim 1}(u) = \gamma_{\phi^{2},q \sim 1}(u) - \gamma_{\phi,q \sim 1}(u)$. Then we can employ the relations $\eta_{q \sim 1}\equiv\gamma_{\phi,q \sim 1}(u_{q \sim 1}^{*})$ and $\nu_{q \sim 1}^{-1}\equiv 2 - \eta_{q \sim 1} - \overline{\gamma}_{\phi^{2},q \sim 1}(u_{q \sim 1}^{*})$ for obtaining the high energy nonextensive critical exponents $\eta_{q \sim 1}$ and $\nu_{q \sim 1}$. Although the high energy nonextensive $\beta_{q \sim 1}$-function, anomalous dimensions and nontrivial fixed point depend explicitly on the nonextensive parameter $q$, this dependence disappears in the high energy nonextensive critical exponents values computation procedure, where the integral $I$ is canceled out in the middle of the  calculations. Then we obtain that the high energy nonextensive critical exponents values, at least up to NLO, are the same as their low energy extensive counterparts \cite{Wilson197475}. This result shows, at least at the loop level just approached, that the nonextensivity of the theory at high or intermediate energies ($q \sim 1$) is broken down, \emph{i. e.} that it is not strong enough to yield critical exponents values depending on the nonextensive $q$ parameter. As argued in Ref \cite{e19010021}, it is difficult to ascertain if the nonextensivity of the theory is violated or not when one approaches $q$-values close to unity. Precisely, this is the aim of the present work. The result above shows that the nonextensivity violation mechanism at intermediate energies probed here is naive and not a true one.

\subsection{Unconventional minimal subtraction scheme}
     
\par The present method is characterized by its generality and elegance \cite{J.Math.Phys.542013093301}, where now the external momenta of Feynman diagrams are not held at a particular fixed value but are free to assume any arbitrary values. The minimal set of needed Feynman diagrams to be computed, up to the next-leading-order, are the ones \cite{Amit}
\begin{eqnarray} 
\parbox{12mm}{\includegraphics[scale=1.0]{fig10.eps}}, \parbox{12mm}{\includegraphics[scale=1.0]{fig6.eps}}, \parbox{12mm}{\includegraphics[scale=.9]{fig7.eps}}, \parbox{12mm}{\includegraphics[scale=.9]{fig21.eps}}
\end{eqnarray}
respectively. The corresponding evaluated diagrams have the following expressions
\begin{eqnarray}
\parbox{10mm}{\includegraphics[scale=1.0]{fig10.eps}} = \frac{1}{\epsilon} \left[1 - \frac{1}{2}\epsilon - \frac{1}{2}\epsilon L_{q \sim 1}(P^{2}, m_{B}^{2}, q) \right],
\end{eqnarray}   
\begin{eqnarray}
&&\parbox{12mm}{\includegraphics[scale=1.0]{fig6.eps}} = \left\{-\frac{3 m_{B}^{2}}{2 \epsilon^{2}}\left[1 + \frac{1}{2}\epsilon + \left(\frac{\pi^{2}}{12} +1 \right)\epsilon^{2} \right] -  \frac{3 m_{B}^{2}}{4}\tilde{i}_{q \sim 1}(P^{2}, m_{B}^{2}, q)  - \right.  \nonumber \\  &&\left. \frac{P^{2}}{8 \epsilon}\left[1 + \frac{1}{4}\epsilon - 2 \epsilon L_{3,q \sim 1}(P^{2}, m_{B}^{2}, q)\right]\right\}, 
\end{eqnarray}
\begin{eqnarray}
&&\parbox{12mm}{\includegraphics[scale=1.0]{fig7.eps}} = \left\{-\frac{5 m_{B}^{2}}{3 \epsilon^{3}}\left[1 + \epsilon + \left(\frac{\pi^{2}}{24} + \frac{15}{4} \right)\epsilon^{2} \right] -  \frac{5 m_{B}^{2}}{2 \epsilon}\tilde{i}_{q \sim 1}(P^{2}, m_{B}^{2}, q) - \right.  \nonumber \\  &&\left.  \frac{P^{2}}{6 \epsilon^{2}}\left[1+ \frac{1}{2}\epsilon - 3 \epsilon L_{3, q \sim 1}(P^{2} + K_{\mu\nu}P^{\mu}P^{\nu},m_{B}^{2})\right]\right\},\quad
\end{eqnarray}
\begin{eqnarray}
\parbox{14mm}{\includegraphics[scale=1.0]{fig21.eps}} = \frac{1}{\epsilon^{2}} \left[1 - \frac{1}{2}\epsilon - \epsilon L_{q \sim 1}(P^{2}, m_{B}^{2}, q) \right],
\end{eqnarray}  
where
\begin{eqnarray}
L_{q \sim 1}(P^{2}, m_{B}^{2}, q) = \int_{0}^{1}dx\ln[x(1-x)P^{2} + qm_{B}^{2}],
\end{eqnarray}
\begin{eqnarray}
L_{3, q \sim 1}(P^{2}, m_{B}^{2}, q) = \int_{0}^{1}dx(1-x) \ln[x(1-x)P^{2} + qm_{B}^{2}],
\end{eqnarray}
\begin{eqnarray}
&&\tilde{i}_{q \sim 1}(P^{2}, m_{B}^{2}, q) = \int_{0}^{1} dx \int_{0}^{1}dy  \ln y \frac{d}{dy}\left((1-y)\times \right.  \nonumber \\  &&\left. ln\left\{y(1-y)P^{2} + \left[1-y + \frac{y}{x(1-x)}\right]qm_{B}^{2} \right\}\right).
\end{eqnarray}
Now, we are in a position to evaluate the high energy nonextensive $\beta_{q \sim 1}$-function, anomalous dimensions and nontrivial fixed point. We then obtain
\begin{eqnarray}\label{rygyfggfgfgyt}
\beta_{q \sim 1}(u) = -\epsilon u +   \frac{N + 8}{6}u^{2} -  \frac{3N + 14}{12}u^{3}, 
\end{eqnarray}
\begin{eqnarray}\label{hkfdusdrs}
\gamma_{\phi,q \sim 1}(u) = \frac{N + 2}{72}u^{2} - \frac{(N + 2)(N + 8)}{1728}u^{3},  
\end{eqnarray}
\begin{eqnarray}\label{sfdtsvdb}
\overline{\gamma}_{\phi^{2}, q \sim 1}(u) = \frac{N + 2}{6}u -  \frac{N + 2}{12}u^{2},
\end{eqnarray}
\begin{eqnarray}
u_{q \sim 1}^{*} = \frac{6\epsilon}{(N + 8)}\Bigg\{ 1 +  \epsilon\Bigg[ \frac{3(3N + 14)}{(N + 8)^{2}} \Bigg]\Bigg\}.
\end{eqnarray}
We can observe that the $q$- and momentum-dependent integrals $L_{q \sim 1}(P^{2}, m_{B}^{2}, q)$, $L_{3, q \sim 1}(P^{2}, m_{B}^{2}, q)$ and $\tilde{i}_{q \sim 1}(P^{2}, m_{B}^{2}, q)$ have been cancelled out and do not contribute to the aforementioned results, as the referred method demands \cite{J.Math.Phys.542013093301}. This fact has implied in the disappearance of a possible dependence of the critical exponents on the nonextensive parameter $q$, thus making the high energy nonextensive results identical to their low energy extensive counterparts. This means that the high energy nonextensive critical exponents, when we apply once again the relations $\eta_{q \sim 1}\equiv\gamma_{\phi,q \sim 1}(u_{q \sim 1}^{*})$ and $\nu_{q \sim 1}^{-1}\equiv 2 - \eta_{q \sim 1} - \overline{\gamma}_{\phi^{2},q \sim 1}(u_{q \sim 1}^{*})$, are the same as the corresponding low energy extensive ones. This result obtained in the current method agrees with the one obtained through the method displayed in the earlier Sect.. This fact confirms the universality of the critical exponents since they are the same when obtained through different and independent methods and shows the arbitrariness of the field theoretical renormalization group method employed as it is required by general renormalization group theory considerations \cite{Wilson197475}.

\subsection{Bogoliubov-Parasyuk-Hepp-Zimmermann method}

\par In the method approached now, we can not compute just a minimal set of Feynman diagrams up to NLO as in the earlier methods. On the other hand, we must compute fourteen diagrams and counterterms \cite{BogoliubovParasyuk,Hepp,Zimmermann}. The present method is known as the BPHZ (Bogoliubov-Parasyuk-Hepp-Zimmermann) method. The theory is renormalized, up to NLO, by the following renormalization constants \cite{Kleinert}
\begin{eqnarray}\label{Zphi}
&& Z_{\phi, q \sim 1}(u,\epsilon^{-1}) = 1 + \frac{1}{P^{2}} \Biggl[ \frac{1}{6} \mathcal{K} 
\left(\parbox{12mm}{\includegraphics[scale=1.0]{fig6.eps}}
\right) \Biggr|_{m^2 = 0} S_{\parbox{10mm}{\includegraphics[scale=0.5]{fig6.eps}}} + \frac{1}{4} \mathcal{K} 
\left(\parbox{12mm}{\includegraphics[scale=1.0]{fig7.eps}} \right) \Biggr|_{m^2 = 0} S_{\parbox{6mm}{\includegraphics[scale=0.5]{fig7.eps}}} + \nonumber \\ && \frac{1}{3} \mathcal{K}
  \left(\parbox{12mm}{\includegraphics[scale=1.0]{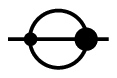}} \right) S_{\parbox{6mm}{\includegraphics[scale=0.5]{fig26.eps}}} \Biggr], 
\end{eqnarray}

\begin{eqnarray}\label{Zg}
&& Z_{u, q \sim 1}(u,\epsilon^{-1}) = 1 + \frac{1}{\mu^{\epsilon}u} \Biggl[ \frac{1}{2} \mathcal{K} 
\left(\parbox{10mm}{\includegraphics[scale=1.0]{fig10.eps}} + 2 \hspace{1mm} perm.
\right) S_{\parbox{10mm}{\includegraphics[scale=0.5]{fig10.eps}}} + \nonumber \\ &&  \frac{1}{4} \mathcal{K} 
\left(\parbox{17mm}{\includegraphics[scale=1.0]{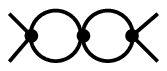}} + 2 \hspace{1mm} perm. \right) S_{\parbox{10mm}{\includegraphics[scale=0.5]{fig11.eps}}} + \frac{1}{2} \mathcal{K} 
\left(\parbox{12mm}{\includegraphics[scale=.8]{fig21.eps}} + 5 \hspace{1mm} perm. \right) S_{\parbox{10mm}{\includegraphics[scale=0.4]{fig21.eps}}} + \nonumber \\ && \frac{1}{2} \mathcal{K} 
\left(\parbox{10mm}{\includegraphics[scale=1.0]{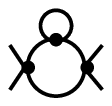}} + 2 \hspace{1mm} perm.
\right) S_{\parbox{10mm}{\includegraphics[scale=0.5]{fig13.eps}}} + \mathcal{K}
  \left(\parbox{10mm}{\includegraphics[scale=1.0]{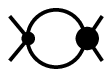}} + 2 \hspace{1mm} perm. \right) S_{\parbox{6mm}{\includegraphics[scale=0.5]{fig25.eps}}} + \nonumber \\ && \mathcal{K}
  \left(\parbox{10mm}{\includegraphics[scale=1.0]{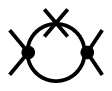}} + 2 \hspace{1mm} perm. \right) S_{\parbox{6mm}{\includegraphics[scale=0.5]{fig24.eps}}} \Biggr],
\end{eqnarray}

\begin{eqnarray}\label{Zphi}
&& Z_{m^{2}, q \sim 1}(u,\epsilon^{-1}) = 1 + \frac{1}{qm^{2}} \Biggl[ \frac{1}{2} \mathcal{K} 
\left(\parbox{12mm}{\includegraphics[scale=1.0]{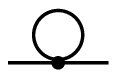}}
\right) S_{\parbox{6mm}{\includegraphics[scale=0.5]{fig1.eps}}} + \frac{1}{4} \mathcal{K} 
\left(\parbox{12mm}{\includegraphics[scale=1.0]{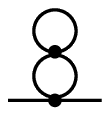}} \right) S_{\parbox{6mm}{\includegraphics[scale=0.5]{fig2.eps}}} + \frac{1}{2} \mathcal{K}
  \left(\parbox{12mm}{\includegraphics[scale=1.0]{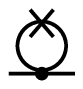}} \right) S_{\parbox{4mm}{\includegraphics[scale=0.5]{fig22.eps}}} + \nonumber \\ && \frac{1}{2} \mathcal{K}
  \left(\parbox{12mm}{\includegraphics[scale=1.0]{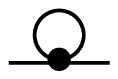}} \right) S_{\parbox{6mm}{\includegraphics[scale=0.5]{fig23.eps}}} + \frac{1}{6} \mathcal{K}
  \left(\parbox{12mm}{\includegraphics[scale=1.0]{fig6.eps}} \right)\Biggr|_{P^{2} = 0} S_{\parbox{6mm}{\includegraphics[scale=0.5]{fig6.eps}}} \Biggr]. \hspace{4mm}
\end{eqnarray}
The symbol $S_{\parbox{6mm}{\includegraphics[scale=0.5]{fig6.eps}}}$ means the symmetry factor for the corresponding diagram for some $N$-component field. When we compute the diagrams and counterterms, we obtain the following results
\begin{eqnarray}
\left(\parbox{12mm}{\includegraphics[scale=1.0]{fig6.eps}}\right)\Biggr|_{m^{2}=0} =  -\frac{u^{2}P^2}{8\epsilon} \left[ 1 + \frac{1}{4}\epsilon -2\epsilon\, J_{3, q \sim 1}(P^{2}, m^{2}, q, \mu) \right],
\end{eqnarray}
\begin{eqnarray}
\parbox{12mm}{\includegraphics[scale=1.0]{fig7.eps}}\bigg|_{m^{2}=0} =  \left[1 + \frac{1}{2}\epsilon - 3\epsilon\, J_{3, q \sim 1}(P^{2}, m^{2}, q, \mu)\right],
\end{eqnarray}
\begin{eqnarray}
\parbox{10mm}{\includegraphics[scale=1.0]{fig26.eps}} \quad = -\frac{3P^{2}u^{3}}{16\epsilon^{2}}\left[1 + \frac{1}{4}\epsilon - 2\epsilon\, J_{3, q \sim 1}(P^{2}, m^{2}, q, \mu)\right],
\end{eqnarray}
\begin{eqnarray}
\parbox{10mm}{\includegraphics[scale=1.0]{fig10.eps}} = \frac{\mu^{\epsilon}u^{2}}{\epsilon} \left[1 - \frac{1}{2}\epsilon - \frac{1}{2}\epsilon J_{q \sim 1}(P^{2}, m^{2}, q, \mu) \right],
\end{eqnarray}
\begin{eqnarray}
\parbox{16mm}{\includegraphics[scale=1.0]{fig11.eps}} = -\frac{\mu^{\epsilon}u^{3}}{\epsilon^{2}} \left[1 - \epsilon - \epsilon J_{q \sim 1}(P^{2}, m^{2}, q, \mu) \right],
\end{eqnarray}
\begin{eqnarray}
\parbox{12mm}{\includegraphics[scale=0.8]{fig21.eps}} = -\frac{\mu^{\epsilon}u^{3}}{2\epsilon^{2}} \left[1 - \frac{1}{2}\epsilon - \epsilon J_{q \sim 1}(P^{2}, m^{2}, q, \mu) \right],
\end{eqnarray}
\begin{eqnarray}
\parbox{12mm}{\includegraphics[scale=1.0]{fig13.eps}} =   \frac{\mu^{\epsilon}u^{3}}{2\epsilon^{2}} J_{4, q \sim 1}(P^{2}, m^{2}, q, \mu),
\end{eqnarray}
\begin{eqnarray}
\parbox{10mm}{\includegraphics[scale=1.0]{fig25.eps}} = \frac{3\mu^{\epsilon}u^{3}}{2\epsilon^{2}} \left[1 - \frac{1}{2}\epsilon - \frac{1}{2}\epsilon J_{q \sim 1}(P^{2}, m^{2}, q, \mu) \right],
\end{eqnarray}
\begin{eqnarray}
\parbox{12mm}{\includegraphics[scale=1.0]{fig24.eps}} =  -\frac{\mu^{\epsilon}u^{3}}{2\epsilon^{2}} J_{4, q \sim 1}(P^{2}, m^{2}, q, \mu),
\end{eqnarray}
\begin{eqnarray}
\parbox{12mm}{\includegraphics[scale=1.0]{fig1.eps}} =
\frac{qm^{2}u}{(4\pi)^{2}\epsilon}\left[ 1 - \frac{1}{2}\epsilon\ln\left(\frac{qm^{2}}{4\pi\mu^{2}}\right)\right],
\end{eqnarray}
\begin{eqnarray}
\parbox{8mm}{\includegraphics[scale=1.0]{fig2.eps}} = - \frac{qm^{2}u^{2}}{(4\pi)^{4}\epsilon^{2}}\left[ 1 - \frac{1}{2}\epsilon - \epsilon\ln\left(\frac{qm^{2}}{4\pi\mu^{2}}\right)\right],
\end{eqnarray}
\begin{eqnarray}
\parbox{12mm}{\includegraphics[scale=1.0]{fig22.eps}} =  \frac{qm^{2}g^{2}}{2\epsilon^{2}}\left[ 1 - \frac{1}{2}\epsilon - \frac{1}{2} \epsilon\ln\left(\frac{qm^{2}}{4\pi\mu^{2}}\right)\right],
\end{eqnarray}
\begin{eqnarray}
\parbox{12mm}{\includegraphics[scale=1.0]{fig23.eps}} =  \frac{3qm^{2}u^{2}}{2\epsilon^{2}}\left[ 1 - \frac{1}{2} \epsilon\ln\left(\frac{qm^{2}}{4\pi\mu^{2}}\right)\right],
\end{eqnarray}
\begin{eqnarray}
\left(\parbox{12mm}{\includegraphics[scale=1.0]{fig6.eps}}\right)\Biggr|_{P^{2} = 0} =  -\frac{3qm^{2}g^{2}}{2\epsilon}\left[ 1 + \frac{1}{2}\epsilon -\epsilon\ln\left(\frac{qm^{2}}{4\pi\mu^{2}}\right)\right],
\end{eqnarray}
where
\begin{eqnarray}\label{uhduhufgjg}
J_{q \sim 1}(P^{2}, m^{2}, q, \mu) = \int_{0}^{1}d x \ln \left[\frac{x(1-x)P^{2} + qm^{2}}{\mu^{2}}\right],\quad
\end{eqnarray}
\begin{eqnarray}\label{uhduhufgjgdhg}
J_{3,q \sim 1}(P^{2}, m^{2}, q, \mu) = \int_{0}^{1}\int_{0}^{1} d x \,d y\,(1-y) \ln \Biggl\{\frac{y(1-y)P^{2}}{\mu^{2}} + \left[1-y + \frac{y}{x(1-x)}  \right]\frac{qm^{2}}{\mu^{2}}\Biggr\},\quad\quad
\end{eqnarray}
\begin{eqnarray}\label{ugujjgdhg}
J_{4,q \sim 1}(P^{2}, m^{2}, q, \mu) = \frac{m^{2}}{\mu^{2}}\int_{0}^{1}d x\frac{(1 - x)}{\frac{x(1 - x)P^{2}}{\mu^{2}} + \frac{qm^{2}}{\mu^{2}}}.
\end{eqnarray}
Thus we can compute the $\beta_{q \sim 1}$-function, anomalous dimensions and nontrivial fixed point and obtain 
\begin{eqnarray}\label{rygyfggffgyt}
\beta_{q \sim 1}(u) = -\epsilon u + \frac{N + 8}{6}u^{2} -  \frac{3N + 14}{12}u^{3}, 
\end{eqnarray}
\begin{eqnarray}\label{hkfusdrs}
\gamma_{\phi,q \sim 1}(u) = \frac{N + 2}{72}u^{2} - \frac{(N + 2)(N + 8)}{1728}u^{3},  
\end{eqnarray}
\begin{eqnarray}\label{kujyhghsghju}
\gamma_{m^{2}, q \sim 1}(u) = \frac{N + 2}{6}u -  \frac{5(N + )2}{72}u^{2},
\end{eqnarray}
\begin{eqnarray}
u_{q \sim 1}^{*} = \frac{6\epsilon}{(N + 8)}\Bigg\{ 1 +  \epsilon\Bigg[ \frac{3(3N + 14)}{(N + 8)^{2}} \Bigg]\Bigg\}.
\end{eqnarray}
Now by applying the relations $\eta_{q \sim 1}\equiv \gamma_{\phi,q \sim 1}(u_{q \sim 1}^{*})$ and $\nu_{q \sim 1}^{-1}\equiv 2 - \gamma_{m^{2},q \sim 1}(u_{q \sim 1}^{*})$, once again we obtain the high energy nonextensive critical exponents values are the same as the low energy extensive ones, although the Feynman diagrams and counterterms depend on the nonextensive parameter $q$ through the the $q$- and momentum-dependent $J_{q \sim 1}(P^{2}, m^{2}, q, \mu)$, $J_{3,q \sim 1}(P^{2}, m^{2}, q, \mu)$ and $J_{4,q \sim 1}(P^{2}, m^{2}, q, \mu)$ integrals. These integrals have disappeared as expected in this method \cite{BogoliubovParasyuk,Hepp,Zimmermann}. This completes our task of computing the high energy nonextensive critical exponents up to NLO for massive theories. In particular, in four dimensions, from the $q$-independence in a physically relevant setup of the renormalization group scheme where the physics is taking place, nothing is left and the corresponding theory is independent of $q$ as can be seen in the expressions for the high energy nonextensive $\beta_{q \sim 1}$-functions and anomalous dimensions computed in the all three distinct renormalization group schemes, \emph{i. e.} that these functions turn out to be the same as their extensive counterparts in four dimensions. Now we have to evaluate the high energy nonextensive critical exponents through massless theories.

\section{high energy nonextensive critical exponents: massless theories}

\par As in massless theories the mass is null and the nonextensivity of the high energy nonextensive theory is explicitly given just by its effective nonextensive mass, then the high energy nonextensive bare free inverse propagator is give by $k^{2}$. This propagator is the same as the one for the low energy extensive massless theory. Thus, the Feynman diagrams needed for the computation of the high energy nonextensive critical exponents in the referred high energy nonextensive massless theories are the same as the ones for the low energy extensive theories. Then the high energy nonextensive critical exponents we obtain though such theories in any of the three methods for massless theories, namely the normalization conditions \cite{Amit}, minimal subtraction scheme \cite{Amit} and the massless BPHZ \cite{Amit} methods, are the same as their low energy extensive counterparts. We then obtain the same results as the ones obtained through the three earlier methods in massive theories. 

\section{high energy nonextensive amplitude ratios}

\par As the complete set of amplitude ratios can be computed only in massless theories with spontaneously symmetry breaking though three distinct and independent methods, namely the normalization conditions \cite{Amit}, minimal subtraction scheme \cite{Amit} and the massless BPHZ \cite{Amit} methods (see the Ref. \cite{Neto2017} for the computation of amplitude ratios, although for Lorentz-violating theories, from which we can recover the low energy extensive theory by taking the limit in which the Lorentz-violating mechanism is vanishing, through these three methods), the same arguments presented in the earlier Sec. can be applied as well. Thus we obtain that the high energy nonextensive amplitude ratios are the same as their low energy extensive counterparts.

\section{Generalization for any loop level}

\par Now we have to generalize the earlier finite NLO results to the high energy nonextensive critical exponents and amplitude ratios for any loop orders. We begin our journey by considering firstly the former universal quantities. Since they are universal, we can choose any of the six renormalization group methods employed here. We have to choose the most general one, \emph{i. e.} the BPHZ method. As a given Feynman diagram in the low energy extensive theory, for some arbitrary loop order, is given by $\mathcal{F}(u,P,m,\epsilon,\mu)$, its high energy nonextensive counterpart can be expressed as $\mathcal{F}_{q \sim 1}(u,P,m,q,\epsilon,\mu) \equiv \mathcal{F}(u,P,qm,\epsilon,\mu)$ through the substitution $m \rightarrow qm$. As in the BPHZ method we have the canceling of any mass-dependent terms for all-loop order in perturbation theory \cite{BogoliubovParasyuk,Hepp,Zimmermann} and the $\beta$-function and anomalous dimensions do not depend on the mass of the theory, we have that the nonextensive parameter $q$ disappears in the middle of calculations through the term $qm$ for all-loop order. Thus, the resulting $\beta_{q \sim 1}$-function and anomalous dimensions for all-loop order are the same as their low energy extensive counterparts
\begin{eqnarray}\label{uhgufhduhufdhu}
\beta_{q \sim 1}(u) =  -\epsilon u + \sum_{n=2}^{\infty}\beta_{n}^{(0)}u^{n}\equiv\beta(u), 
\end{eqnarray}
\begin{eqnarray}
\gamma_{\phi,q \sim 1}(u) = \sum_{n=2}^{\infty}\gamma_{n}^{(0)}u^{n}\equiv\gamma_{\phi}(u),
\end{eqnarray}
\begin{eqnarray}
\gamma_{m^{2},q \sim 1}(u) = \sum_{n=1}^{\infty}\gamma_{m^{2},n}^{(0)}u^{n}\equiv\gamma_{m^{2}}(u),
\end{eqnarray}
where $\beta_{n}^{(0)}$, $\gamma_{\phi,n}^{(0)}$ and $\gamma_{m^{2},n}^{(0)}$ are the $n$-th loop radiative quantum corrections to the corresponding low energy extensive functions. Now, if we compute the high energy nonextensive nontrivial fixed point $u_{q \sim 1}^{\*}$ valid for all-loop level, which is obtained as the nontrivial solution for the high energy nonextensive condition $\beta_{q \sim 1}(u_{q \sim 1}^{*}) = 0$ for the high energy nonextensive $\beta_{q \sim 1}$-function valid for all-loop level, we will find that it is the same as their all-loop order low energy extensive counterpart. Then, the corresponding all-loop order high energy nonextensive critical exponents are the same as their all-loop low energy extensive counterparts. This completes our generalization for any loop level. Then, the assertion that the nonextensivity of the theory at high or intermediate energies ($q \sim 1$) is broken down or is not strong enough to yield $q$-dependent critical exponents is now valid for all loop orders. The next step is to approach the earlier task but now for amplitude ratios. This task is attained by applying the same arguments used in the earlier Sec. for NLO but now for any loop level. As the theory used for computing high energy nonextensive amplitude ratios valid for all-loop order is some all-loop level massless one, the resulting theory does not depend on the nonextensive parameter $q$, since a possible $q$-dependence of the theory could come only through a massive term of the form $qm$ which is vanishing for a massless theory. This completes our task. We can now proceed to present our conclusions.

\section{Conclusions}\label{Conclusions}

\par We evaluated the all-loop radiative quantum corrections to the universal critical exponents and amplitude ratios for O($N$) $\lambda\phi^{4}$ scalar high energy nonextensive $q$-field theories. For attaining that goal, we generalized the results for that universal quantities obtained here for finite NLO through six distinct and independent methods for the high energy nonextensive critical exponents and three different and independent methods for the amplitude ratios. The results for these universal quantities obtained through that distinct methods were all identical among them thus showing the arbitrariness of the field-theoretic renormalization group method employed and the possibility of checking the results by comparing the results obtained through so many distinct and independent methods. Furthermore, the high energy nonextensive critical exponents and amplitude ratios were the same as their extensive counterparts. This fact shows that the nonextensivity of the theory is broken down at high or intermediate energies, \emph{i. e.} that the nonextensivity property, which is present at very high energies, is not strong enough to survive at high or intermediate energies. The present work opens a new research branch since we can approach similar investigations by computing corrections to scaling and finite-size scaling effects for critical exponents as well as for amplitude ratios at the high or intermediate energy domain.

\bibliography{apstemplate}

\end{document}